\documentclass[referee]{aa} 
\usepackage{graphicx}
\usepackage{txfonts}
\begin{document}
   \title{Quasi-Hilda Comet 147P/Kushida-Muramatsu}

   \subtitle{Another long temporary satellite capture by Jupiter}

   \author{K. Ohtsuka \inst{1},
          T. Ito \inst{2},
          M. Yoshikawa \inst{3},
          D. J. Asher \inst{4},
           \and
          H. Arakida \inst{5}
          }

   \offprints{K. Ohtsuka}

   \institute{Tokyo Meteor Network, Daisawa 1--27--5, Setagaya-ku, Tokyo 155--0032, Japan\\
              \email{ohtsuka@jb3.so-net.ne.jp}
         \and
             National Astronomical Observatory of Japan,
             Osawa 2--21--1, Mitaka, Tokyo 181--8588, Japan%
         \and
             JAXA, Yoshinodai 3--1--1, Sagamihara, 
             Kanagawa 229--8510, Japan%
         \and
             Armagh Observatory, College Hill, Armagh, BT61 9DG, UK
         \and
             Waseda University, Nishi-Waseda 1--6--1, Shinjuku-ku,
             Tokyo 169--8050, Japan \\
             }

   \date{Received ; accepted }

 
  \abstract
   {The quasi-Hilda comets (QHCs), 
   being in unstable 3:2 Jovian mean motion 
   resonance, are considered a major cause 
   of temporary satellite capture (TSC) by Jupiter. 
   Though the QHCs may be escaped Hilda asteroids, 
   their origin and nature have not yet been studied in sufficient detail. 
   Of particular interest are long TSCs/orbiters. 
   Orbiters -- in which at least one full revolution about the 
   planet is completed -- are rare astronomical events; 
   only four have been known to occur in the last several decades. 
   Every case has been associated with 
   a QHC: 82P/Gehrels 3; 111P/Helin-Roman-Crockett; 
   P/1996 R2 (Lagerkvist); and the possibly QHC-derived 
   D/1993 F2 (Shoemaker-Levy 9, SL9). 
   }
   {We focus on long TSC/orbiter events involving QHCs and Jupiter. 
   Thus we survey the known QHCs, searching for further long TSCs/orbiters 
   over the past century. 
   }
   {First, we confirmed the long TSC/orbiter 
   events of 82P, 111P, and 1996 R2 in order to 
   test our method against previous work, 
   applying a general $N$-body Newtonian code. 
   We then used the same procedure to survey the remaining known QHCs 
   and search for long TSC/orbiter events.
   }
   {We newly identified another long TSC/orbiter: 
   147P/Kushida-Muramatsu from 1949 May 
   14$^{+97 {\rm days}}_{-106 {\rm days}}$--1961 July 15. 
   Our result is verified by integrations of 243 cloned orbits which 
   take account of the present orbital uncertainty of this comet. The 
   event involves an $L_{\rm 2} \to L_{\rm 1}$ transition as with 
   82P and 1996 R2; this may represent a distinct subtype of TSCs from 
   QHC derived ($L_{\rm 1} \to$) longer captures 
   exemplified by 111P and (probably) SL9, 
   though this classification is still only based on a small database of TSCs. 
   }
   {This is the third long TSC and the fifth orbiter to be found, 
   thus long TSC/orbiter events involving Jupiter have 
   occurred once per decade. 
   Two full revolutions about Jupiter were completed 
   and the capture duration was $12.17^{+0.29}_{-0.27}$ 
   years; both these numbers rank 147P as third among long TSC/orbiter 
   events, behind SL9 and 111P. This study also confirms the importance of 
   the QHC region as a dynamical route into and out of Jovian TSC, 
   via the Hill's sphere. 
   }


   \authorrunning{Ohtsuka et al.}
   \titlerunning{Quasi-Hilda comet 147P/Kushida-Muramatsu}
   \maketitle
%

\section{Introduction}

   Among all the asteroids that have been recorded up to the present 
   in the asteroid database, e.g., the ``JPL Small-Body Database" 
   ({\tt http://ssd.jpl.nasa.gov/}), a large number (more than 1000 
   including unnumbered objects) 
   are known to populate the region of the 3:2 mean motion 
   resonance (MMR) with Jupiter, in the outer main belt. 
   These are the ``Hilda asteroids" (Schubart \cite{schubart68}, 
   \cite{schubart82}, \cite{schubart}; 
   Ip \cite{ip}; Yoshikawa \cite{yoshikawa}; 
   Franklin et al. \cite{franklin}; 
   Nesvorn\'y \& Ferraz-Mello \cite{nesvorny}). 
   Their semimajor axes, $a$, concentrate in the range 
   3.7 AU $\le a \le$ 4.2 AU, at eccentricities $e \le 0.3$, 
   and inclinations $i \le 20^\circ$ (Zellner et al. \cite{zellner}). 
   This results in a range of the Tisserand parameter with 
   respect to Jupiter, $T_{\rm J}$, of $\sim 2.90$--3.05, where 
   $T_{\rm J} = a_{\rm J}/a + 
   2 \sqrt {a/a_{\rm J}(1 - e^2)} \cos I$, 
   with $a_{\rm J}$ being the semimajor axis 
   of Jupiter and $I$ the mutual inclination between the 
   orbits. 
   The critical arguments for the Hildas, 
   $\phi = 3\lambda_{\rm J}-2\lambda-\varpi$, librate about $0^\circ$, 
   being stable in the long-term ($\lambda$ is mean longitude, 
   $\varpi$ is longitude of perihelion, and J indicates Jupiter). 
   As regards physical properties, the low-albedo D- and P-types  
   are more abundant in Hildas' surface colours than the small 
   fraction of C-types (Dahlgren \& Lagerkvist \cite{dahlgren}; 
   Dahlgren et al. \cite{dahlgren97}; 
   Gil-Hutton \& Brunini \cite{gil-hutton}; 
   Licandro et al. \cite{licandro}). 
   The surface colour of D- and P-type 
   asteroids, such as Hildas and Trojans in the outer main 
   belt, corresponds well with that of cometary nuclei 
   (Fitzsimmons et al. \cite{fitzsimmons}; Jewitt \cite{jewitt02}), 
   which means that they are covered with 
   a similar mineralogical surface, suggestive of a common origin.
  
   More than 50 Jovian irregular satellites are known at present 
   (Jewitt \& Haghighipour \cite{jewitt}). 
   \'Cuk \& Burns (\cite{cuk}) pointed out that 
   the progenitor of the main prograde cluster, the Himalia family, 
   was plausibly derived from the Hildas long ago, 
   if it was captured by a gas-drag assisted mechanism. 
   Thus tracing the origin and nature of irregular satellites 
   is very significant for studying the accretion processes 
   in the early solar system. 
   Satellite capture mechanics in the circular 
   restricted three-body problem (CR3BP) 
   or the $N$-body problem  
   has often been investigated 
   (e.g., H\'enon \cite{henon}; 
   Huang \& Innanen \cite{huang}; Tanikawa \cite{tanikawa}; 
   Murison \cite{murison}; Brunini et al. \cite{brunini}; 
   Nesvorn\'y et al. \cite{nesvorny03}, \cite{nesvorny07}).  
   Reflectance spectra of Jovian irregulars, 
   being dominated by D- and C-types 
   (Luu \cite{luu}; Grav et al. \cite{grav}), 
   are comparable to those of Hildas. 
   
   During the past half century, several Jupiter family comets 
   (JFCs; cf.\ Levison \cite{levison}) 
   have stayed in or near the Hilda zone, 
   although being in unstable 3:2 MMR with Jupiter. 
   Some of them have been transferred from outside 
   to inside, \textit{vice versa}, or from inside to inside of 
   Jupiter's orbit by undergoing a temporary satellite 
   capture (TSC) by Jupiter (e.g., Carusi \& Valsecchi 
   \cite{carusi79}; Tancredi et al. \cite{tancredi}). 
   Such a JFC is called a ``quasi-Hilda comet" (QHC) 
   by Kres\'ak (\cite{kresak}), who identified three such objects: 
   39P/Oterma, 74P/Smirnova-Chernykh, and 82P/Gehrels 3. 
   Di Sisto et al. (\cite{di sisto}) 
   integrated the motions of 500 fictitious Hildas for 
   $\sim 10^9$ years, 
   and found that most of them escaped from the Hilda zone 
   into the JFC population, i.e., left the 3:2 MMR 
   and evolved quickly on to unstable orbits: such a chaotic 
   diffusion from the Hilda zone has also been demonstrated 
   by Nesvorn\'y \& Ferraz-Mello (\cite{nesvorny}). 
   In addition, large-scale collisional processes, 
   such as the late heavy bombardment, 
   might also release small bodies from 
   the Hilda zone into JFCs, e.g., see 
   Gil-Hutton \& Brunini (\cite{gil-hutton00}). 
   Hence, some QHCs may indeed be 
   such escaped Hildas themselves. 
   Recently, Toth (\cite{toth}) updated the QHC list, 
   finding a total of 17 members (see Section 3). 
   This includes bodies such as the
   renowned Comet D/1993 F2 (Shoemaker-Levy 9, SL9) 
   that have undergone TSC by Jupiter and then disappeared after
   colliding with the planet. 
   Surface spectroscopic (or colorimetric) measurements for 
   QHCs have only been carried out for 82P 
   (De Sanctis et al. \cite{de sanctis}), the results 
   of which also indicate a taxonomic D-type.  

   Among the QHCs, 39P/Oterma was the first known to be 
   temporarily captured by Jupiter, in 1936--1938 
   (Marsden \cite{marsden}). 
   However, this comet flew through the region near Jupiter 
   over a rather short time, during which the comet did not complete 
   a full revolution orbiting about the planet. 
   In contrast, unlike 39P's ``fly-through" capture, 
   there is a different kind of TSC, in which at least 
   one full revolution about the planet is completed; we 
   deal with these in the present paper. 
   Following Kary \& Dones (\cite{kary}) we call such objects
   ``orbiters". These are often characterized 
   by a long capture with very small perijove distance, 
   usually lasting for $\sim 10$ years or more. Not 
   all orbiters become such long TSCs ($> 10$ yr), although 
   of course they last longer than the fly-through type.  
   SL9 is a representative case for both long TSCs and orbiters. 
   This comet was pointed out to have possibly been 
   QHC-derived before its tidal disruption on passing 
   through perijove at less than 1.5 Jovian equatorial radii 
   ($R_{\rm J}$, where $R_{\rm J}=71492.4$ km), 
   i.e., within the Roche limit, in 1992 July and 
   its subsequent collision with Jupiter in 1994 July 
   (Nakano \& Marsden \cite{nakano93}; Sitarski \cite{sitarski}; 
   Benner \& McKinnon \cite{benner}). 
   If it is QHC-derived, then SL9 is the only 
   QHC so far that has been orbiting the planet as a TSC 
   at the time of discovery (Benner \& McKinnon \cite{benner}). 
   By numerically integrating SL9's 
   pre-collision orbital motion, 
   several studies showed that 
   the TSC duration of SL9 lasted for 50 years or more, during which 
   the comet completed more than 30 revolutions orbiting about Jupiter, 
   making it nominally the longest known TSC (Carusi et al. \cite{carusi}; 
   Benner \& McKinnon \cite{benner}; 
   Chodas \& Yeomans \cite{chodas}). 
   However, according to Benner \& McKinnon (\cite{benner}), 
   SL9 was the most chaotic known object in the solar system 
   with an effective Lyapunov time of only $\sim10$ years on its
   jovicentric orbit. 
   Thus it is difficult to determine with certainty 
   SL9's pre-capture orbit and its true TSC duration. 
   Nevertheless, Benner \& McKinnon's additional statistical analysis 
   of distributions in $a$-$e$ space and $T_{\rm J}$ values, 
   based on back integrations of the various SL9 fragments, revealed a 
   possible QHC origin of SL9. 
   The work of Kary \& Dones (\cite{kary}) supports this possibility: 
   they traced the motions of numerous fictitious 
   JFCs for $\sim 10^5$ years 
   and found that half the SL9-like 
   very long captures $>50$ years were due to QHCs. 
   They estimated that the frequency of such a very long 
   TSC (= ``long capture" as designated by them) 
   is extremely rare, only 0.02\% 
   of all the TSC events in their simulations. 
   Interestingly, impacts on Jupiter are more frequent 
   than the very long TSCs by a factor of 8--9. 
   They also evaluated that long TSCs (= ``orbiters bound $> 10$ yr" as 
   designated by them) and orbiters are still rare events, 
   at the respective levels of 0.8\% 
   ($\supset$ very long TSCs) and 2\% ($\supset$ long TSCs) 
   relative to all events, with about 98\% being the 
   short TSC type which contains the fly-through events. 
   Carusi \& Valsecchi (\cite{carusi79}) had earlier 
   confirmed the rarity of orbiter events, 
   simulating motions of fictitious small bodies as well.

   In the jovicentric Keplerian system, 
   a TSC (especially a long TSC/orbiter) occurs 
   whenever a small body passes near one of the collinear 
   libration points 
   $L_{\rm 1}$ or $L_{\rm 2}$ in the CR3BP of 
   the Sun-Jupiter-(third) body system 
   with very low velocity, 
   i.e., effectively becoming bound by Jupiter when it enters 
   the Hill's region with near-zero velocity. 
   After that, the bound small body revolves about 
   Jupiter on an elliptical jovicentric orbit until it again 
   passes through the region near either $L_{\rm 1}$ or $L_{\rm 2}$ 
   and escapes from the Jovian system. 
   Considering such a transition using newly developed 
   dynamical systems techniques based on a Hamiltonian formulation 
   in the CR3BP, Koon et al. (\cite{koon}) and 
   Howell et al. (\cite{howell}) demonstrated that a TSC by Jupiter 
   occurs when the small body passes through a region inside the 
   invariant manifold structure related to periodic halo 
   orbits around $L_{\rm 1}$ or $L_{\rm 2}$ in the Hill's region. 
   The TSC (or its duration) is usually defined by the 
   jovicentric Kepler energy, $E_{\rm J}$, being negative, 
   $E_{\rm J} < 0$, with the additional 
   condition that the bound small body 
   must be within the jovicentric sphere of gravitational influence: 
   Kary \& Dones (\cite{kary}) set its boundary 
   at 3 Hill's sphere radii (=1.065 AU) of Jupiter. 
   However, Howell et al. (\cite{howell}) defined the TSC 
   duration as the residence time in the Hill's region. 
   The former generally lasts longer than the latter, 
   and here we regard the former as the TSC duration. 
   The dynamics involved in TSC is quite different 
   from that of quasi-satellites in 1:1 libration 
   with Jupiter far outside the Hill's region 
   (Wiegert et al. \cite{wiegert}; 
   Kinoshita \& Nakai \cite{kinoshita}).

   Apart from SL9, only three orbiters
   have been known to occur. Every case has been associated 
   with a QHC: 82P; 111P/Helin-Roman-Crockett; 
   and, though it is not in Toth's (\cite{toth}) QHC list, 
   P/1996 R2 (Lagerkvist).
   These TSCs were found by 
   Rickman (\cite{rickman}), Tancredi et al. (\cite{tancredi}), 
   and Hahn \& Lagerkvist (\cite{hahn}), respectively, and are 
   discussed further in Section 2. 
   The occurrence of these few events 
   during the last several decades is consistent with the rarity of 
   long TSC/orbiter events suggested by Kary \& Dones (\cite{kary}). 
   Although 74P encountered Jupiter at distances of 0.24 AU and 0.47 AU 
   in 1955 October and 1963 September respectively, it 
   was not bound to Jupiter (Rickman \cite{rickman}; 
   Carusi et al. \cite{carusi85b}); this comet is, 
   however, expected to experience a TSC by Jupiter 
   in this century (Carusi et al. \cite{carusi85b} and see also Section 4).  
   There further exist some JFCs that 
   encountered Jupiter more closely than several QHCs involved in TSC, 
   e.g., 16P/Brooks 2; D/1770 L1 (Lexell); 81P/Wild 2;
   but they were not captured by the planet owing to 
   their high-velocity encounters (Carusi et al. \cite{carusi85a}; 
   Emel'yanenko \cite{emel'yanenko}). 

   Therefore studying the origin and nature of QHCs is of great importance 
   from various astronomical points of view mentioned above, especially 
   as regards their origin and being possibly related to the Hildas 
   and the Jovian irregular satellites. 
   Such studies may provide unique knowledge and clues about 
   formation processes in the early solar system. 
   Here we focus on long TSC/orbiter events involving Jupiter and QHCs. 
   First (Section 2), applying a general $N$-body Newtonian code, 
   we reconfirmed the long TSC/orbiter 
   events of 82P, 111P, and 1996 R2. 
   Then (Section 3) we used the same procedure to search 
   Toth's (\cite{toth}) QHCs list for other objects that have become 
   long TSCs/orbiters in the past century. 
   Eventually, we successfully found another long TSC/orbiter, 
   occurring in the mid-20th century, 147P/Kushida-Muramatsu.


\section{Computation method and its application to the known TSCs}

   The jovicentric trajectories for the long TSCs/orbiters, 
   except for SL9, have been numerically simulated by 
   Carusi \& Valsecchi (\cite{carusi79}) and 
   Carusi et al. (\cite{carusi85a}, \cite{carusi85b}) 
   for 82P, by Tancredi et al. (\cite{tancredi}), 
   Belbruno \& Marsden (\cite{belbruno}) 
   and Howell et al. (\cite{howell}) for 111P, 
   and by Hahn \& Lagerkvist (\cite{hahn}) for 1996 R2. 

   First of all, we attempted to reproduce their 
   TSC events in order to compare our simulations 
   with these previous studies, 
   integrating back to the time of each TSC. 
   We applied a general $N$-body Newtonian code, 
   then equations of motion for the $i$th body are:
   \begin{equation}
    \frac{d^2 \mathbf {r}_i}{dt^2}= 
    - G \frac{(M_{\mbox{\sun}} + m_i)
    \mathbf {r}_i}{{r_i}^3}
    -{\sum_{j=1; j \neq i}^{N}} G m_j
    \left (\frac{\mathbf{r}_i - \mathbf{r}_j}
    {{r_{ij}}^3} + \frac{\mathbf{r}_j}
    {{r_j}^3} \right),    
   \end{equation}
   where $G$ is the gravitational constant, 
   $M_{\mbox{\sun}}$ and $m_i$ are respectively 
   the mass of the sun and the $i$th body ($i=1, 2, \cdots, N$), 
   $\mathbf {r}_i$ is the heliocentric 
   position vector of the $i$th body, 
   $r_i$ is the heliocentric distance of the 
   $i$th body, and $r_{ij}$ is the distance between 
   the $i$th and $j$th bodies. Here, 
   we regarded comets as massless bodies.
   
   The integrator that we applied here is the 
   ``SOLEX", Ver. 9.1 package, developed by 
   Vitagliano (\cite{vitagliano}) based
   on the Bulirsh-Stoer method. 
   Coordinates and velocities of the planets, regarded as point masses, 
   were based on the JPL Planetary and Lunar Ephemeris DE409. 
   We confirmed that the results of our 
   numerical integrations did not significantly change 
   when we used other integration methods that 
   we have often applied in our studies, e.g., 
   the Adams method. 
   Our integrator can accurately process very close encounters 
   by means of a routine that makes automatic time step adjustments, 
   and truncation and round-off errors 
   are almost negligible for our investigation here. 
   Therefore the SOLEX integrator is 
   sufficiently reliable to deal with the problem of 
   close encounters with Jupiter.

   As initial parameters, up to date osculating orbital elements were 
   taken from the JPL Small-Body Database, mentioned above in Section 1, 
   for 82P and 1996 R2 and from Nakano (\cite{nakano}) for 111P, 
   as listed in Table~\ref{tbl:QHCorbits}. 
   82P and 111P are numbered, multiple-apparition comets, 
   covered by very long observational arcs, hence their orbital solutions 
   are highly precise. Meanwhile, although 1996 R2 is unnumbered and still a 
   one-apparition comet, consequently with a shorter arc, 
   the number of astrometric positions is the most, 135, 
   and the RMS residual is the least, $0''.76$, among the three comets, 
   hence we judged that the TSC of 1996 R2 is worth simulating here. 
   The nongravitational parameters for these QHCs' motions were 
   not detectable from their astrometry hence were not 
   included in our integrations -- in any case
   their motions are chaotic
   during TSC. 
   The Jovian oblateness terms and 
   the perturbations by the Galilean 
   satellites were also ignored since their effects are 
   negligible in our investigations 
   (see Kary \& Dones \cite{kary}); 
   these effects should be included in analyses of 
   very long TSCs of SL9-like objects having highly inclined 
   Jupiter-grazing orbits.

\begin{figure*}
   \centering
   \includegraphics[width=17cm]{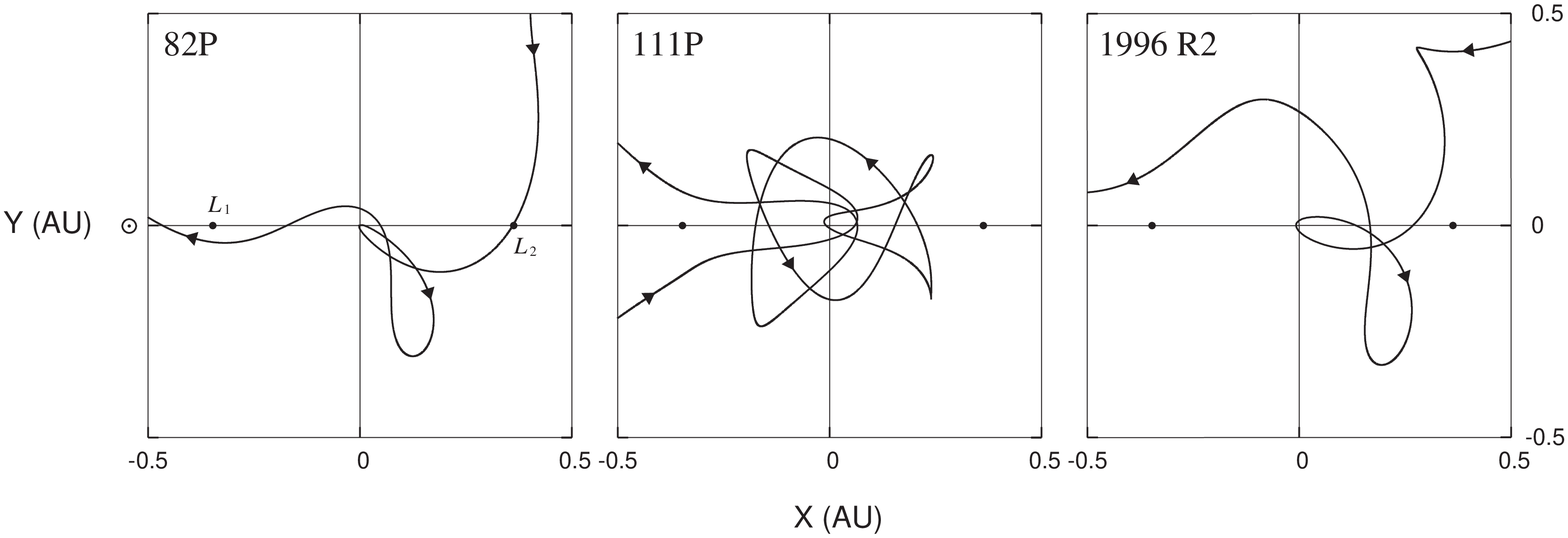}
   \caption{Simulated TSC trajectories 
               for 82P, 111P, and 1996 R2.
               Plots show $x$-$y$ projections, over 1 AU square, 
               onto the heliocentric orbital plane of Jupiter, 
               in the jovicentric rotating frame where 
               the sun is always in the direction of the $-x$-axis. The arrows 
               indicate the direction of time (opposite to the direction of
               the integrations, which went backwards in time).
               }
              \label{F1}%
   \end{figure*}

   Our simulated TSC trajectories of 82P, 111P, 
   and 1996 R2 are shown in Figure~\ref{F1}. 
   In these plots we use a rotating frame in which the jovicentric 
   rectangular coordinates are referred to the heliocentric 
   orbital plane of Jupiter, with the sun located 
   in the direction of the $-x$-axis, 
   and the $+z$-axis pointing to the north.
   In Figure~\ref{F1} every TSC motion is unstable and chaotic, 
   with substantial solar perturbations occurring near each
   comet's apojove.

   Among these TSCs, the trajectories of 82P and 1996 R2 are 
   similar in profile to each other. 
   Both these QHCs came tangentially from outside Jupiter's orbit 
   with a low-inclination retrograde motion, 
   i.e., their pre-capture heliocentric 
   orbits imply that they came from the Centaur region, their perihelia being 
   beyond the Jovian orbit. 
   During the TSC, first they passed close to $L_{\rm 2}$ with low 
   relative velocities (between Jupiter and comets in the jovicentric 
   frame) 
   of respectively $\sim 0.7$ km s$^{-1}$ and 
   $\sim 0.6$ km s$^{-1}$, then revolved about Jupiter, 
   completing one full revolution, and finally escaped from 
   the Jovian system into the QHC region, 
   passing near $L_{\rm 1}$.
   In the $L_{\rm 2} \to L_{\rm 1}$ transition, 
   an object that enters the Jovian system close to 
   $L_{\rm 2}$ seems to 
   have enough orbital energy to rapidly escape through 
   the region near $L_{\rm 1}$ (Tanikawa \cite{tanikawa}). 
   The long lasting captures for 82P and 1996 R2 may be somewhat 
   related to the orbital stability for the retrograde satellites 
   (Huang \& Innanen \cite{huang}; 
   Nesvorn\'y et al. \cite{nesvorny03}). However, 
   the TSC duration is below 10 years in both cases; so to be more exact
   these orbiters should perhaps not be categorized as long TSCs. 

   On the other hand, the motion of 111P was 
   more complicated than those of 82P and 1996 R2. 
   111P completed three full revolutions about 
   Jupiter during its long TSC of 18.45 years, 
   both of which numbers rank second next to SL9 among known TSCs. 
   Its pre-capture and post-escape heliocentric orbits 
   were both QHC type (Belbruno \& Marsden \cite{belbruno}), 
   entering through the region near $L_{\rm 1}$ at 
   a relative velocity of  
   $\sim 1.0$ km s$^{-1}$ and escaping near $L_{\rm 1}$ again. 
   Its capture/escape trajectory is almost symmetric about 
   the $x$-axis (cf.\ Murison \cite{murison}) 
   inside the Hill's region in the CR3BP. 
   Howell et al. (\cite{howell}) demonstrated that the 111P 
   capture/escape trajectory is almost identical to 
   some asymptotic trajectory winding onto a periodic halo 
   orbit of $L_{\rm 1}$, such trajectories  
   forming segments of the invariant manifolds.

   We therefore reconfirmed the long TSC/orbiter events 
   involving 82P, 111P and 1996 R2. 
   Our orbiter trajectories look exactly as in 
   Belbruno \& Marsden (\cite{belbruno}) 
   and Howell et al. (\cite{howell}) for 111P, 
   and as in all the above papers for 82P, 
   but the results by Tancredi et al. (\cite{tancredi}) for 
   111P and Hahn \& Lagerkvist (\cite{hahn}) for 1996 R2 show 
   somewhat different behaviours. 
   Both these can be ascribed to slight differences in 
   initial parameters -- 
   their initial orbits were
   based on observational arcs of respectively 2--4 months 
   and 111 days whereas ours were based on longer time spans 
   of 4643 and 153 days. 
   Besides, Hahn \& Lagerkvist (\cite{hahn}) plotted 
   the trajectory not of the nominal orbit but of 
   a clone, i.e., a possible TSC trajectory  
   within the uncertainty of the initial parameters. 
   In fact, we could not reach the TSC of 1996 R2 
   by using their nominal orbit. 
   This demonstrates that in studying TSC events, 
   initial parameters must be determined precisely,
   based on long-arc improved orbital solutions 
   wherever possible, 
   since jovicentric orbits during TSC have 
   a chaotic nature.

   The capture situations of these QHCs are summarized in 
   Table~\ref{tbl:TSC}. 
   During these TSCs, 82P, 111P and 1996 R2 
   experienced 2, 6 and 3 perijove passages, respectively. 
   In particular, 82P passed its closest perijove of 
   $3.01 R_{\rm J}$ in 1970 August, 
   just outside the Roche limit for comets 
   $\sim 2.7 R_{\rm J}$ (assuming a cometary bulk density 
   $\rho$ $\sim 1$ g cm$^{-3}$). 
   Table~\ref{tbl:TSC} also shows the minimum values of the 
   jovicentric Kepler energy $E_{\rm J}=-1/2a'$ (in AU$^{-1}$) 
   where $a'$ is the semimajor axis for the jovicentric orbit. 
   The low values attained, $-2.80$, $-3.20$, and $-3.28$ for the three
   comets, are often a feature of long TSCs/orbiters, as against $-1.79$ 
   which we have computed for 39P during its fly-through in the 1930s. 

   Shortly after TSC, these three QHCs were discovered and 
   observed as ``quasi-Hilda comets" for the first time. 
   The time difference, $\Delta T_{\rm {(TSC-obs)}}$, 
   between the end time of the TSC, given in Table~\ref{tbl:TSC}, 
   and the first astrometric time, given in Table~\ref{tbl:QHCorbits}, 
   is notable.
   Every $\Delta T_{\rm {(TSC-obs)}}$ is short enough 
   (1.3 years for 82P, 3.49 years for 111P, and 3.16 years for 1996 R2) 
   to indicate that the simulated TSC should be a real event. 

\section{147P/Kushida-Muramatsu: another long TSC/orbiter}

   Next we surveyed the remaining objects in Toth's (\cite{toth}) 
   QHC list (Table~\ref{tbl:QHC list}) to determine 
   whether other long TSCs/orbiters around Jupiter exist. 
   We applied our code to integrate each orbit 
   back 100 years from its initial epoch, 
   since the Lyapunov time of JFCs is rather short, 
   generally 100 $\pm$ 50 yr (Tancredi \cite{tancredi95}). 
   Moreover, going back more than 100 yr may not be meaningful 
   because of likely weak nongravitational accelerations, 
   in addition to the accumulation of orbital errors. 
   We recognized that short TSC events have sometimes occurred, 
   for which results will be published 
   as an additional paper elsewhere. 
   Eventually, we successfully found another long TSC, 
   involving 147P/Kushida-Muramatsu.

   The observational history of 147P is described by 
   Green (\cite{green93}, \cite{green}) and 
   Kronk (\cite{kronk}). 
   The comet was photographically discovered by Y. Kushida 
   and O. Muramatsu (Yatsugatake South Base Observatory,  
   Japan) on 1993 December 8.65, 
   using a 0.25-m f/3.4 reflector. 
   It was then at magnitude $\sim16.5$ 
   and slightly diffuse, about $1$--$2'$ in diameter 
   with a central condensation. 
   The comet was recovered on 2000 October 3.72, when 
   T. Oribe (Saji Observatory, Japan) obtained 
   CCD images with the 1.03-m reflector. 
   Thus 147P has been detected at two perihelion returns, 
   allowing it to be numbered, although it has not
   been astrometrically observed since 2002 March.

\begin{figure*}
   \centering
   \includegraphics[width=12cm]{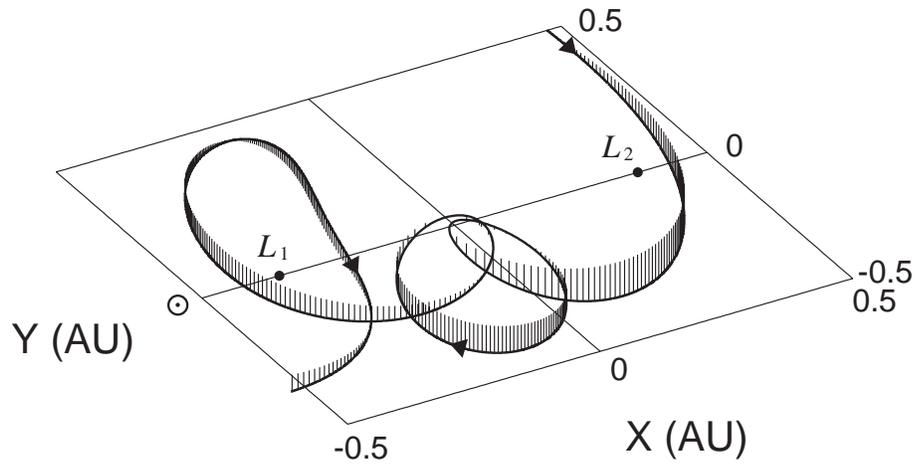}
   \caption{3D-view of 147P's trajectory during TSC  
               in the jovicentric rotating frame covering 1 AU square. 
               The vertical lines, spaced at 10-day intervals, each 
               connect a position of the comet on the trajectory with 
               its projection onto the $x$-$y$ plane.
               }
              \label{F2}%
   \end{figure*}

   In our integration of 147P, the initial parameters 
   were taken from Nakano (\cite{nakano02}), as listed in 
   Table~\ref{tbl:147Porbit}. 
   His orbital solution was accurately improved based on
   197 carefully selected positions covering an arc of 
   3024 days and also provided associated error estimates 
   (Table~\ref{tbl:147Porbit}). 
   We believe that this solution is the best determined for 
   this comet, probably based on a longer arc and more
   astrometric positions than any other.
   The nongravitational accelerations in 147P's motion were 
   not detected from the astrometry.
   Integrating back this orbit,
   we found that the comet was captured by Jupiter 
   in the mid-20th century. The 3D-view of 
   147P's TSC situation is illustrated in Figure~\ref{F2}.

   However, 
   $\Delta T_{\rm {(TSC-obs)}}$ of 147P is extremely large, 
   more than 30 years, which greatly exceeds the values of 
   $< 3.5$ years for 82P, 111P and 1996 R2 (and also SL9 
   with negative $\Delta T_{\rm {(TSC-obs)}}$). 
   Thus we must further examine whether or not 
   147P really underwent a long TSC/orbiter phase 
   in the mid-20th century. 
   To answer this, we traced 
   other possible orbital motions of 147P back to and during its TSC, 
   generating multiple ``clones" at initial epoch, 
   and integrating them. 
   As well as the nominal 147P 
   osculating orbit, other clones 
   had slightly different orbital elements, 
   within the $\pm 1 \sigma$ error, given in Table~\ref{tbl:147Porbit}. 
   We generated the clones on the basis of 
   three possible values [nominal and 
   $\pm 1\sigma$] for five 
   orbital elements [$a$; $e$; $\omega$; 
   $\mathit{\Omega}$; $i$]. 
   This number of permutations yielded a total of 243 $(=3^5)$ clones, 
   including the nominal one. 
   We also generated some test clones varying 
   T (or $M$) within the allowed error, confirming that this effect is 
   negligible; thus we excluded this parameter 
   when generating the main set of clones. 
   Conversely, we found $a$ to be the most strongly effective parameter 
   in causing divergence among the clones' orbital behaviours. 
   
\begin{figure*}
   \centering
   \includegraphics[width=12cm]{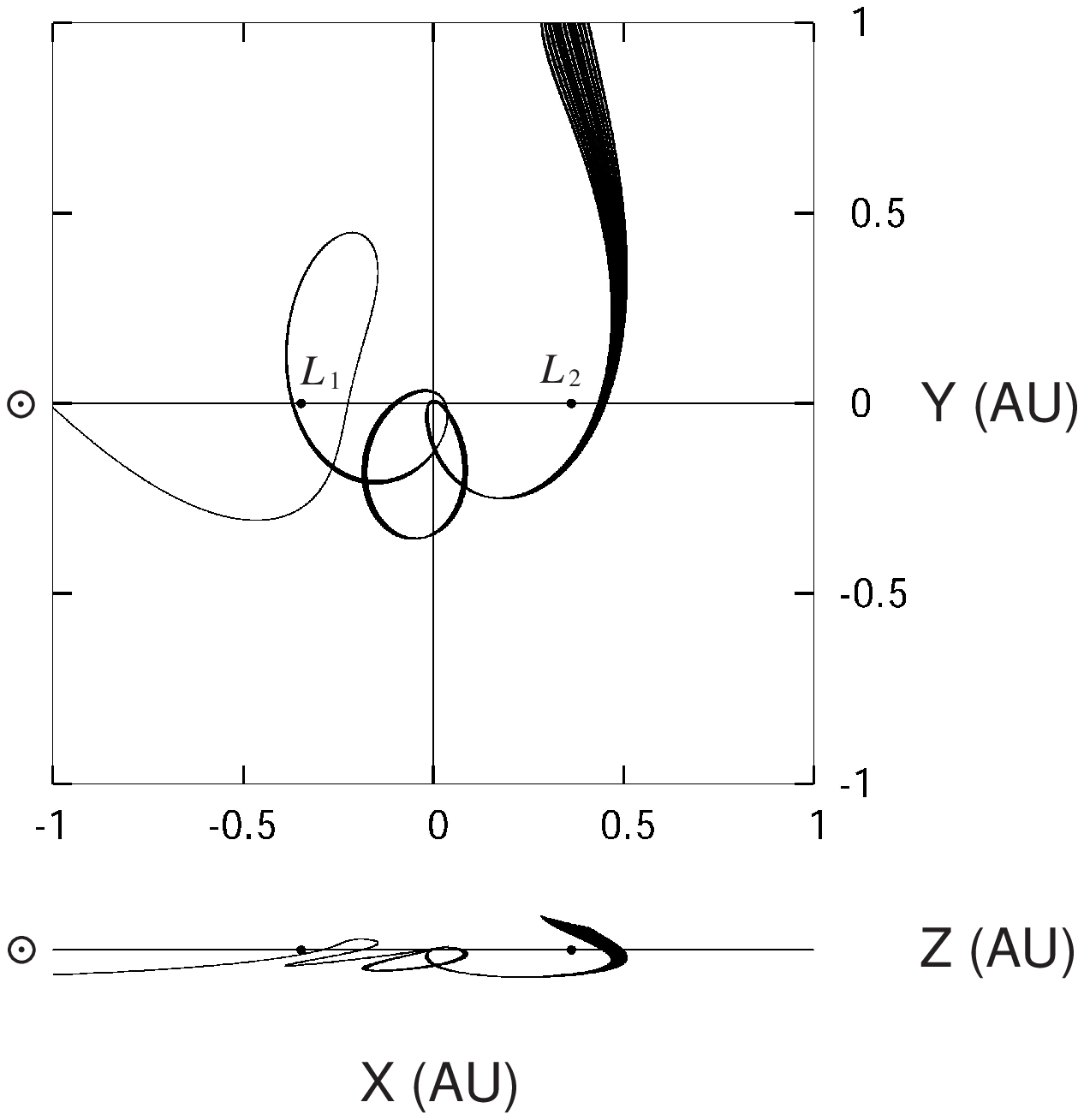}
   \caption{$x$-$y$ and $x$-$z$ projections of TSC trajectories for 243 clones
               generated from 147P. They underwent long TSCs, 
               completing two full revolutions about Jupiter during 
               the $L_{\rm 2} \to L_{\rm 1}$ transition. 
               The dispersion of their trajectories corresponds to 
               $\pm 1 \sigma$ error in the orbit 
               determination. 
               Hence we conclude that the real 147P indeed experienced 
               a long TSC/orbiter event in the mid-20th century.
               }
              \label{F3}%
   \end{figure*}

   We found that all 243 clones underwent TSCs.
   In Figure~\ref{F3}, we can see 
   that their trajectories become scattered in their pre-capture orbital 
   phase, while converging toward the post-escape phase, 
   since the orbital motions were integrated back in time. 
   However, the dispersion of their TSC 
   trajectories, corresponding to $\pm 1 \sigma$ error in the orbit 
   determination, is not as scattered as is sometimes 
   expected with chaotic trajectories. 
   We conclude that it is very likely that 147P really was an 
   orbiter, experiencing a long TSC in the mid-20th century.
   The motions of all the clones follow 
   almost the same profile as those of 82P and 1996 R2, 
   coming tangentially from outside Jupiter's orbit, 
   i.e., with a low-inclination retrograde motion in the jovicentric frame.
   The pre-capture heliocentric Centaur orbit of 147P had $a \sim 6.2$ AU, 
   $q \sim 5.8$ AU, and $e \sim 0.07$.
   The comet entered the Jovian system passing through the region 
   near $L_{\rm 2}$ with relative velocity 
   $\sim 0.9$ km s$^{-1}$, revolved about Jupiter, 
   escaped from the Jovian system passing near $L_{\rm 1}$, 
   and was discovered and observed as a QHC after $\sim 32$ years.
   However, an important difference from 
   both 82P and 1996 R2 is that 147P completed two full 
   revolutions about Jupiter, thus experiencing a significantly longer TSC.
      
\begin{figure*}
   \centering
   \includegraphics{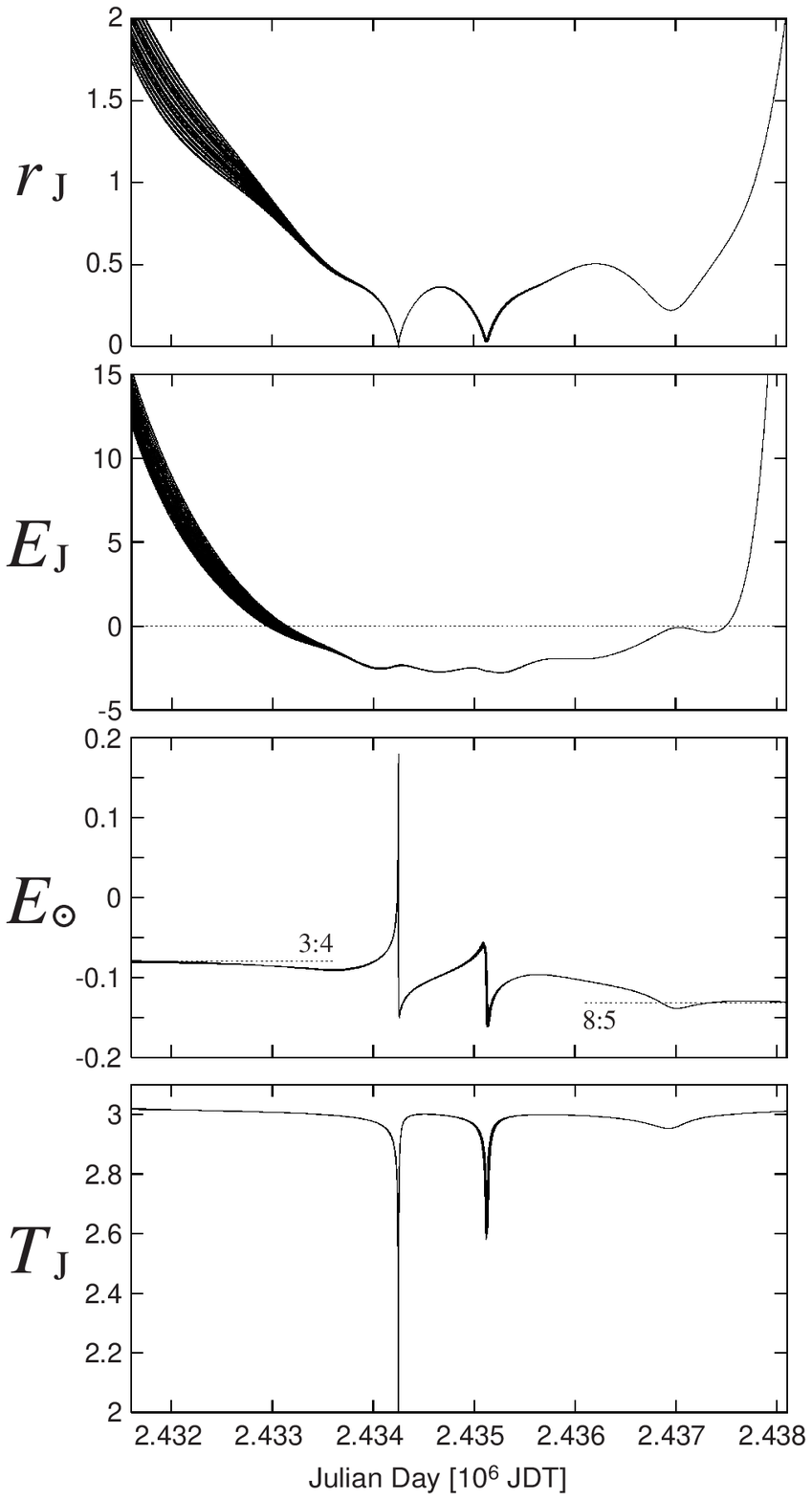}
   \caption{ Variations of $r_{\rm J}, E_{\rm J}, 
               E_{\mbox{\sun}}$ and $T_{\rm J}$ 
               of 243 clones of 147P 
               in and around their interval of TSC, where 
               $r_{\rm J}=$ jovicentric distance (in AU); 
               $E_{\rm J}$ and $E_{\mbox{\sun}}=$ jovicentric 
               and heliocentric Kepler energy, respectively; 
               $T_{\rm J}=$ Tisserand parameter. 
               The abscissa for all the plots covers 
               2431596.5 (1945 May 21.0 TT) to 
               2438095.5 (1963 March 07.0 TT), 
               the time for which the nominal 147P 
               was within 2 AU of Jupiter.
               }
              \label{F4}%
    \end{figure*}

   Figure~\ref{F4} shows the variations in
   $r_{\rm J}, E_{\rm J}, E_{\mbox{\sun}}$, and $T_{\rm J}$ 
   of all the clones in and around the TSC interval, 
   where $r_{\rm J}$ is jovicentric distance (in AU), 
   and $E_{\mbox{\sun}}$ is heliocentric 
   Kepler energy $= -1/2a$ (in AU$^{-1}$). 
   The $r_{\rm J}$ and $E_{\rm J}$ diagrams 
   in the pre-capture orbital phase display a scatter that is as evident 
   as the scatter in Figure~\ref{F3}.  
   We can see (most clearly from the $r_{\rm J}$ diagram, but with 
   signs also reflected in the other three plots) that 147P 
   experienced three perijove passages.
   The term of negative $E_{\rm J}$ corresponds to TSC, 
   during which $r_{\rm J}$ was always within 3 Hill's sphere radii,  
   as defined in Section 1; thus 147P underwent a 
   long TSC for $12.17^{+0.29}_{-0.27}$ years (where error 
   estimates are based on the clones' dispersion). 
   The minimum $E_{\rm J}$ taking a low value of $-2.78 \pm 0.01$ 
   is also an expected feature of a long TSC/orbiter event.
   In the $E_{\mbox{\sun}}$ diagram, 
   we can see the sharp spike 
   corresponding to the closest encounter to Jupiter, 
   during which the nominal heliocentric orbit is briefly hyperbolic.
   The value of $T_{\rm J}$ was slightly higher than 3.0 
   in the pre-capture orbital phase, 
   as is often associated with low-velocity encounters 
   at Jupiter (Kres\'ak \cite{kresak}), and it suddenly dropped 
   twice around the closest perijove passages. 
   We summarize the capture situation data of 147P in 
   Table~\ref{tbl:147PTSC} (again, error estimates come from the 
   range shown by all the clones).

   Also indicated in the $E_{\mbox{\sun}}$ diagram in Figure~\ref{F4} is 
   the commensurability of 147P with Jupiter changing 
   from 3:4 Centaur-type in the pre-capture orbital phase to 
   8:5 (rather than 3:2) QHC in the post-escape phase. 
   This, however, corresponds to the osculating orbit 
   immediately post-escape, and integrations for motions of 
   all the clones over a longer timescale suggest instead 
   that 147P might librate in the 3:2 Jovian MMR for 
   $\sim 350$ yr following the TSC (and is doing so at the present 
   time), though with a somewhat faster libration period 
   and higher libration amplitude ($\sim 135^\circ$ for the nominal orbit) 
   than the more stable, 
   typical Hilda asteroid orbits  
   (e.g., Schubart \cite{schubart82}, \cite{schubart}; 
   Franklin et al. \cite{franklin}; 
   Nesvorn\'y \& Ferraz-Mello \cite{nesvorny}).
   Moreover, test integrations of the other objects suggest that of the 
   QHCs known to have been involved in a TSC, only 147P has undergone an 
   interval where the critical argument of the 3:2 MMR librates. 
   
   Further backward integrations of 147P inform us of 
   the possibility of another TSC somewhere between 
   the late-19th to early-20th century. 
   However, as all the clones were extremely scattered by that time, 
   we do not consider that TSC here.


\section{Concluding Remarks}

   On the basis of our investigations above, 
   we have presented a newly identified TSC of a comet by Jupiter.
   This is in the rare, orbiter class of TSCs and involves 147P from 
   1949 May 14$^{+97 {\rm days}}_{-106 {\rm days}}$--1961 
   July 15. 
   This is the third long TSC of $>10$ years and 
   the fifth orbiter found, 
   so that Jupiter's long TSCs/orbiters 
   have occurred once per 
   decade. The completion of two full revolutions about Jupiter 
   and the capture duration of $12.17^{+0.29}_{-0.27}$ 
   years rank 147P as third in both these numbers among known orbiters, 
   behind SL9 and 111P. 

\subsection{TSC Classification}

   Following Kary \& Dones (\cite{kary}), 
   we classify the known TSCs as follows: 
   39P as fly-through, 82P and 1996 R2 as orbiters, 
   111P and 147P as long TSCs ($>10$ years), 
   and SL9 as a very long TSC ($>50$ years). 

   On the other hand, depending on the TSC characteristics, 
   Howell et al. (\cite{howell}) defined two TSC types: 
   Type 1 as a 39P-like fly-through; Type 2 as a long lasting capture like
   111P, where the comet experiences more than one close encounter with
   Jupiter while in the TSC region. 
   Here we tentatively divide Type 2 into two subtypes: Type 2A as 
   the $L_{\rm 2} \to L_{\rm 1}$ transition such as 82P, 1996 R2 
   and 147P; Type 2B as the QHC-derived ($L_{\rm 1} \to$) 
   longer capture than 2A, as with 111P and SL9. 
   Further TSC classifications may be possible in the future, 
   if different kinds of TSC are found. 

\subsection{Future TSCs}

   We also surveyed all the known QHCs, integrating their orbital motions 
   forward for 100 years to check for future long TSCs/orbiters. 
   We found that 111P will undergo a long TSC/orbiter phase  
   with 6 perijove passages, and minimum $E_{\rm J} \sim -3.19$, 
   from 2068 April 20--2086 June 09 (duration $\sim 18.14$ years). 
   We identified a rather long capture of 82P 
   with minimum $E_{\rm J} \sim -2.26$ though it is not 
   an orbiter but instead follows a symmetric trajectory about the $x$-axis,  
   from 2056 February 18--2064 July 26 
   (duration $\sim 8.43$ years). 
   There appear to be two TSCs for 74P 
   if we take its initial parameters from 
   the JPL Small-Body Database, 
   from 2025 August 10--2031 May 29 
   (duration $\sim 5.80$ years, minimum 
   $E_{\rm J} \sim -0.74$) and 
   2081 January 29--2085 January 29 
   (duration $\sim 4.00$ years, minimum 
   $E_{\rm J} \sim -0.77$). 

   Such future TSCs have already been 
   predicted by Carusi et al. (\cite{carusi85a}, 
   \cite{carusi85b}) for 74P and 82P and 
   Tancredi et al. (\cite{tancredi}) for 111P, 
   and our simulations are almost identical in 
   profile with those. Meanwhile, 
   it is unlikely that 1996 R2 and 147P will encounter 
   Jupiter as a long TSC/orbiter for 
   the next 100 years from each initial epoch. 
   No further long TSCs/orbiters were found among 
   the remaining QHCs.

\subsection{Tidal Effects}

   Here we discuss the Jovian tidal force 
   acting on 147P and other captured QHCs, 
   around their perijove passages. 
   Well-known cometary tidal splitting events have 
   occurred twice: 16P in 1886 
   (Sekanina \& Yeomans \cite{sekanina}) 
   and SL9, both passing perijove within 
   the Roche limit for comets, $\sim 2.7 R_{\rm J}$. 
   In contrast, the close encounter of 147P with Jupiter
   around 1952 August 26 (Table~\ref{tbl:147PTSC}) was
   at a perijove distance of $14.61^{+2.94}_{-2.61} R_{\rm J}$, 
   comfortably outside the Roche limit.
   The radius, $R_{\rm c}$, of 147P's nucleus was estimated at 0.21 km by 
   Lamy et al. (\cite{lamy}), the smallest 
   among all their measured cometary nuclei. 
   
   The Jovian tidal stress, 
   $\sigma_{\rm T}$, acting on small bodies (QHCs here), 
   is given by: 
   \begin{equation}
    \sigma_{\rm T} \approx Gm_{\rm J}\rho R_{\rm c}^2/r_{\rm J}^3, 
   \end{equation}
   where $m_{\rm J}$ is the mass of Jupiter. 
   Compared to the $\sigma_{\rm T}$ 
   acting on SL9 at perijove $< 1.5 R_{\rm J}$ 
   in 1992 (Scotti \& Melosh \cite{scotti}; 
   Asphaug \& Benz \cite{asphaug94}, \cite{asphaug}), the 
   $\sigma_{\rm T}$ for 147P was extremely small, amounting to $< 0.005 \%$ 
   if we take
   the original, pre-encounter $R_{\rm c}$ of SL9 
   as $\sim 1$ km (Scotti \& Melosh \cite{scotti}; 
   Asphaug \& Benz \cite{asphaug94}) and 
   $\rho$ of both the comets as being equal. Hence, 
   although our treatment here is approximate, there 
   seems no reason to believe that 147P was affected by Jovian tides. 
   On the other hand, 82P, which passed its perijove 
   at $3.01 R_{\rm J}$ in 1970, 
   just outside the cometary Roche limit, 
   may have suffered some effect due to the Jovian tides, 
   even though it was not broken up. 
   Taking $R_{\rm c}$ for 82P as 
   0.73 km (Lamy et al. \cite{lamy}), and equal $\rho$ as above, 
   implies that the $\sigma_{\rm T}$ acting on 82P could be up to 
   $7 \%$ of that for SL9. 
   If 82P is a friable and furthermore a loose rubble-pile 
   object, the heating energy from continual Jovian tides 
   might sufficiently affect
   the comet's nucleus structure so as to allow H$_2$O ice, if present, 
   to sublimate even though 82P is beyond the usual heliocentric 
   distance for H$_2$O sublimation. 
   If our hypothetical scenario is true, 
   such tidal heating effects could trigger 
   cometary activity on 82P and produce an outburst. 
   In this event, the coma size and consequent brightness 
   would peak several days (or more) after the maximum tidal 
   effects. 

   Another intriguing point is whether or not 
   other tidal effects, e.g., tidal distortion and 
   tidal torques leading to rotation state changes 
   (Scheeres et al.\ \cite{scheeres}), 
   are detectable in the light-curve 
   observations of the captured QHCs. 
   In the physical database of cometary nuclei 
   by Lamy et al. (\cite{lamy}), 
   we can notice a rather high axis ratio ($a/b > 1.6$) and 
   long rotation period $\sim 50$ hr in the dataset of 
   82P, though an interpretation in terms of 
   such tidal effects is still speculative.

\subsection{Comets or asteroids?}

   At any rate, we have considered here only a small subset of the TSC 
   events, focusing on the long TSCs/orbiters, 
   for which there still exist only limited available data. 
   Therefore we know little about such unique and 
   extraordinary astronomical events, 
   which are still full of ambiguities. 
   In addition, we do not know whether QHCs are 
   comets or asteroids. Indeed, several QHCs have sometimes been 
   discovered or recovered as asteroids because of their 
   cometary activity being weak: e.g., 36P (= 1925 QD = 1940 RP), 
   39P (= 1950 CR),
   74P (= 1967 EU = 1978 NA$_6$ = 1981 UH$_{18}$ = 1982 YG$_3$), 
   D/1977 C1 (= 1977 DV$_3$), P/1999 XN$_{120}$, P/2001 YX$_{127}$, and
   P/2003 CP$_7$.
   Further observations and research for the 
   QHCs will be necessary to unlock their origin and nature.

\begin{acknowledgements}
      We wish to thank Prof. A. Vitagliano for providing the 
      SOLEX package. We are also grateful to an anonymous 
      reviewer for useful comments that improved the paper. 
      Part of this work was supported by the 2008 
      domestic fellowship of the Astronomical Society of Japan 
      for KO. 
\end{acknowledgements}

\clearpage

\begin{table*}[htb]
\caption{Initial parameters of QHCs 82P/Gehrels 3, 111P/Helin-Roman-Crockett, and P/1996 R2 Lagerkvist (equinox J2000)}
\label{tbl:QHCorbits}
\centering
\begin{tabular}{lccc}
\hline
\hline\noalign{\smallskip}
object & 82P & 111P & 1996 R2 \\%
\hline\noalign{\smallskip}
osculation epoch (TT) & 2002 Mar 27.0 & 2004 Dec 21.0 & 1996 Oct 08.0 \\
mean anomaly $M$      &  $24^\circ.27332$ & $359^\circ.25243$ & $346^\circ.17074$ \\
semimajor axis $a$ (AU) & 4.1388678 & 4.0400597 & 3.7820178 \\
eccentricity $e$ & 0.1238743 & 0.1402634 & 0.3099920 \\
argument of perihelion $\omega$ & $227^\circ.65115$ & $10^\circ.56557$ & $334^\circ.04513$ \\
longitude of ascending node $\mathit{\Omega}$ & $239^\circ.62851$ & $91^\circ.93769$ & $40^\circ.24084$ \\
inclination $i$  & $1^\circ.12654$ & $4^\circ.23300$ & $2^\circ.60532$ \\
number of astrometric positions &  103 & 126 & 135 \\
astrometric arc  & 1975 Oct 27--2002 Mar 08 &  1989 Jan 03--2001 Sep 20 & 1996 Aug 12--1997 Jan 12 \\ [-1pt]  
     & 9629 days (26.36 yr) &  4643 days (12.71 yr) & 153 days (0.42 yr) \\
RMS residual & $0''.79$ & $0''.86$ & $0''.76$ \\
source reference & JPL 12 & Nakano (2005) & JPL 18 \\
\noalign{\smallskip}\hline
\end{tabular}
\end{table*}

\clearpage

\begin{table*}[htb]
\caption{Long TSC/orbiter data of QHCs 82P, 111P, and 1996 R2}
\label{tbl:TSC}
\centering
\begin{tabular}{lccccccc}
\hline
\hline\noalign{\smallskip}
object & TSC duration & no. & min. & heliocentric orbit \&  & perijove time $^{\mathrm{d}}$ & dist.$^{\mathrm{e}}$ & $V_{\rm J}$ $^{\mathrm{f}}$\\ [2pt]
      &  (TT)  & rev.$^{\mathrm{a}}$ & $E_{\rm J}$$^{\mathrm{b}}$ & $L_i$ to $L_j$ transition$^{\mathrm{c}}$   &  (TT) & (AU) & (km s$^{-1}$) \\ [2pt]
\hline\noalign{\smallskip} 
82P & 1966 Dec 11--1974 Jul 11 & 1 & $-2.80$ & $2:3 \to 3:2$ & 1970 Aug 15.6 & 0.0014 (3.01 $R_{\rm J}$) & 34.25  \\
    & 7.58 yr & && $L_{\rm 2} \to L_{\rm 1}$  & 1973 Mar 24.9 & 0.041 (85.16 $R_{\rm J}$) & 9.18 \\
[3pt]
111P & 1967 Jan 25--1985 Jul 08 & 3 & $-3.20$ & $3:2 \to 3:2$ & 1969 Oct 14.4 & 0.057 (119.4 $R_{\rm J}$) & 5.00 \\
    & 18.45 yr & && $L_{\rm 1} \to L_{\rm 1}$ & 1972 Sep 03.2 & 0.175 (336.4 $R_{\rm J}$) & 2.33 \\
    &   &  & && 1976 Apr 11.2 & 0.012 (24.31 $R_{\rm J}$) & 11.85 \\ 
    &   &  & && 1979 Jul 18.6 & 0.211 (442.4 $R_{\rm J}$) & 2.11 \\
    &   &  & && 1980 Nov 17.0 & 0.201 (421.1 $R_{\rm J}$) & 2.20 \\
    &   &  & && 1983 Aug 10.1 & 0.063 (131.7 $R_{\rm J}$) & 4.74 \\
[3pt]
1996 R2 & 1983 Sep 07--1993 Jun 13 & 1 & $-3.28$ & $7:13 \to 8:5$ & 1987 Mar 19.7 & 0.0074 (15.53 $R_{\rm J}$) & 14.92 \\
    & 9.77 yr &  && $L_{\rm 2} \to L_{\rm 1}$ & 1990 Jun 28.8 & 0.169 (352.7 $R_{\rm J}$) & 2.78 \\
    &   &  &  && 1992 Sep 18.5 & 0.325 (680.0 $R_{\rm J}$) & 1.82 \\
\noalign{\smallskip}\hline
\end{tabular}
\begin{list}{}{}
\item[$^{\mathrm{a}}$] number of completed full revolutions about Jupiter
\item[$^{\mathrm{b}}$] minimum $E_{\rm J}$ in AU$^{-1}$ -- occurred on
                       1970 July 30 for 82P, 1976 February 11 for 111P,
                       and 1987 March 15 for 1996 R2
\item[$^{\mathrm{c}}$] Transition associated with the TSC:  firstly, the
                       Jovian MMR (heliocentric orbit) the
                       comet is closest to immediately before/after the TSC; 
                       and secondly, the $L_i$ point near which the comet 
                       passes at the start/end of the TSC phase.
                       
\item[$^{\mathrm{d}}$] time of perijove passage in TT
\item[$^{\mathrm{e}}$] perijove distance in AU and also in $R_{\rm J}$
\item[$^{\mathrm{f}}$] jovicentric velocity in km s$^{-1}$ at perijove
\end{list}
\end{table*}

\clearpage

\begin{table*}[htb]
\caption{Quasi-Hilda comets}
\label{tbl:QHC list}
\centering
\begin{tabular}{llll}
\hline\noalign{\smallskip}
Numbered QHCs: &&& \\ [2pt]
36P/Whipple & 39P/Oterma$^{\mathrm{\#}}$ & 74P/Smirnova-Chernykh & 77P/Longmore \\
82P/Gehrels 3$^{\mathrm{\#}}$ & 111P/Helin-Roman-Crockett$^{\mathrm{\#}}$ & 117P/Helin-Roman-Alu 1 & 129P/Shoemaker-Levy 3 \\
135P/Shoemaker-Levy 8 & 147P/Kushida-Muramatsu & & \\
[3pt]
Unnumbered QHCs: &&& \\ [2pt]
D/1977 C1 (Skiff-Kosai) & D/1993 F2 (Shoemaker-Levy 9)$^{\mathrm{\#}}$ & P/1996 R2 (Lagerkvist)$^{\mathrm{a} \mathrm{\#}}$ & P/1999 XN$_{120}$ (Catalina) \\
P/2001 YX$_{127}$ (LINEAR) & P/2003 CP$_7$ (LINEAR-NEAT) & P/2002 O8 (NEAT) & P/2004 F3 (NEAT) \\
\noalign{\smallskip}\hline
\end{tabular}
\begin{list}{}{}
\item[$^{\mathrm{a}}$] not included in Toth's (\cite{toth}) QHC list
\item[$^{\mathrm{\#}}$] involved in previously known TSC
\end{list}
\end{table*}

\clearpage

\begin{table*}[htb]
\caption{Initial parameters of QHC 147P/Kushida-Muramatsu (equinox J2000) and their $\pm 1 \sigma$ error estimates (Nakano 2002)}
\label{tbl:147Porbit}
\centering
\begin{tabular}{lcc}
\hline\noalign{\smallskip}
osculation epoch (TT) & 2001 May 11.0 & \\
perihelion time T (TT) & 2001 Apr 29.48509 & $\pm 0.00081$ \\
               & ($\Rightarrow M=1^\circ.52643$) & \\
perihelion distance $q$ (AU) & 2.7524408 & $\pm 0.0000019$ \\
semimajor axis $a$ (AU) & 3.8094146 & $\pm 0.0000010$ \\
eccentricity $e$ & 0.2774636 & $\pm 0.0000005$ \\
argument of perihelion $\omega$ & $347^\circ.55482$ & $\pm 0^\circ.00047$ \\
longitude of ascending node $\mathit{\Omega}$ & $93^\circ.69336$ & $\pm 0^\circ.00044$ \\
inclination $i$  & $2^\circ.36694$ & $\pm 0^\circ.00004$ \\
number of astrometric positions &  197 & \\
astrometric arc  & 1993 Dec 08--2002 Mar 20 & \\ [-1pt]
     & 3024 days (8.28 yr) & \\
RMS residual & $0''.85$ & \\
\noalign{\smallskip}\hline
\end{tabular}
\end{table*}

\clearpage

\begin{table*}[htb]
\caption{Long TSC data of QHC 147P}
\label{tbl:147PTSC}
\centering
\begin{tabular*}{17cm}{ccccccc}
\hline
\hline\noalign{\smallskip}
TSC duration & no. & min.& heliocentric orbit \& & perijove time & dist. & $V_{\rm J}$ \\ [2pt]
   (TT)     & rev. &$E_{\rm J}$$^{\mathrm{a}}$&  $L_i$ to $L_j$ transition   &  (TT) & (AU) & (km s$^{-1}$) \\ [2pt]
\hline\noalign{\smallskip}
 1949 May 14--1961 Jul 15  & 2 &$-2.78$& $3:4 \to 8:5$ & 1952 Aug 26.6$^{+1.3}_{-2.2}$ & 0.0070$^{+0.0014}_{-0.0012}$ (14.61$^{+2.94}_{-2.61} R_{\rm J}$) & 15.45$^{+1.62}_{-1.38}$  \\ [2pt]
    ($=$ JDT $2433050.5^{+97}_{-106}$-- & & $\pm 0.01$ & $L_{\rm 2} \to L_{\rm 1}$  & 1955 Jan 18.8$^{+9.8}_{-10.0}$ & 0.0273$^{+0.0013}_{-0.0014}$ (57.05$^{+2.82}_{-2.94} R_{\rm J}$) & 7.59$^{+0.22}_{-0.20}$ \\ [2pt]
     JDT 2437495.5)& &&& 1960 Jan 20.86$\pm 0.02$ & 0.220 (460.60$\pm 0.11 R_{\rm J}$) & 2.73 \\ [2pt]
     $12.17^{+0.29}_{-0.27}$ yr &  & &&& \\
\noalign{\smallskip}\hline
\end{tabular*}
\begin{list}{}{}
\item[$^{\mathrm{a}}$] occurred on 1955 May 28.0  (JDT 2435255.5) $\pm 6$ days 
\end{list}
\end{table*}

\clearpage

\clearpage

\clearpage

\clearpage

\end{document}